\documentclass[reprint,twocolumn,amsmath,amssymb,superscriptaddress]{revtex4-2}

\usepackage{graphicx, graphpap}
\usepackage[dvipsnames]{xcolor}
\usepackage[caption=false,subrefformat=parens,labelformat=parens]{subfig}
\usepackage{amsmath}
\usepackage{overpic}
\usepackage{physics}
\usepackage{enumitem}
\usepackage{comment}
\usepackage{amssymb} 
\usepackage{amsthm}
\usepackage{pstricks}
\usepackage{float}
\usepackage{multirow}
\usepackage{hyperref}
\usepackage{graphicx}
\usepackage[T1]{fontenc}
\usepackage{framed} 
\usepackage{bm}
\usepackage{bbm}

\definecolor{shadecolor}{gray}{0.9}
\definecolor{ECCgreen}{RGB}{0, 153, 0}
\definecolor{EditPurple}{RGB}{160, 32, 240}

\renewcommand{\i}{\mathrm{i}}

\renewcommand{\Tr}[1]{\mathrm{Tr}\left[#1\right]}

\usepackage{placeins}

\begin{document}
	
\title{A Practical Framework for Analyzing High-Dimensional QKD Setups}
\author{Florian Kanitschar}
\email{florian.kanitschar@outlook.com}
\affiliation{Vienna Center for Quantum Science and Technology (VCQ), Atominstitut, Technische Universität Wien, Stadionallee 2, 1020 Vienna, Austria}
\affiliation{AIT  Austrian  Institute  of  Technology,  Center  for  Digital  Safety\&Security,  Giefinggasse  4,  1210  Vienna, Austria}
\author{Marcus Huber}
\affiliation{Vienna Center for Quantum Science and Technology (VCQ), Atominstitut, Technische Universität Wien, Stadionallee 2, 1020 Vienna, Austria}
 \affiliation{Institute for Quantum Optics and Quantum Information (IQOQI),
Austrian Academy of Sciences, Boltzmanngasse 3, 1090 Vienna, Austria}

\date{\today}
	
\begin{abstract}

High-dimensional (HD) entanglement promises both enhanced key rates and overcoming obstacles faced by modern-day quantum communication. However, modern convex optimization-based security arguments are limited by computational constraints; thus, accessible dimensions are far exceeded by progress in HD photonics, bringing forth a need for efficient methods to compute key rates for large encoding dimensions. In response to this problem, we present a flexible analytic framework facilitated by the dual of a semidefinite program and diagonalizing operators inspired by entanglement-witness theory, enabling the efficient computation of key rates in high-dimensional systems. To facilitate the latter, we show how matrix completion techniques can be incorporated to effectively yield improved, computable bounds on the key rate in paradigmatic high-dimensional systems of time- or frequency-bin entangled photons and beyond, revealing the potential for very high dimensions to surpass low-dimensional protocols already with existing technology. In our accompanying work \cite{Kanitschar_2025}, we show how our findings can be used to establish finite-size security against coherent attacks for general HD-QKD protocols both in the fixed- and variable-length scenario.

\end{abstract}
	
\maketitle


\twocolumngrid
\textbf{Introduction.---} Quantum key distribution is one of the most mature quantum technologies, yet it still faces fundamental challenges to be overcome for long-distance applications. While the exponential loss in optical fibers can be overcome by moving to free space, other issues such as low key rate and, in particular, low noise tolerance prevail. The latter limits free-space and satellite-based realizations to nighttime operations, severely reducing uptime and a significant impediment to practical feasibility. Recent works have shown efforts to extend operation times gradually, mainly by experimental adaptations \cite{Han_22, Abasifard_2023, Krzic_2023, Li_2023, Bouchard_2021, Avesani_2021, Liao_2017}. However, this did not turn out sufficiently to close this gap, and it hinted that additional developments from the theoretical and protocol sides are required. 

Elaborate forms of entanglement beyond qubit entanglement are a promising platform to address this issue. High-dimensional (HD) entanglement \cite{Kwiat_1997, Barreiro_2005, Martin_2017,  Islam_2017, Bouchard_2021, Avesani_2021, Bulla_2022, Sulimany_2023}, besides naturally increasing the key rate per signal, has proven to enhance background noise resistance \cite{Ecker_2019} in entanglement-distribution tasks. High-dimensionally entangled states can be produced in labs in various degrees of freedom, among others in the temporal domain \cite{Bavaresco_2018, Hu_2020, Schneeloch_2019}, the frequency domain \cite{Ponce_2022}, and in the form orbital angular momentum entanglement \cite{Maxwell_2022}, as well as in combinations of those, leading to hyperentangled states \cite{Sairam_2022, Yasir_2022}. However, from the theoretical side, the available tools for calculating secure key rates in high-dimensional quantum systems are somewhat limited. On the one hand, there are methods \cite{Sheridan_2010, Doda_2021} requiring Alice and Bob to measure between two and $d+1$ mutually unbiased bases, which is already practically infeasible in low to medium dimensions. On the other hand, numerical methods \cite{Araujo_2022, BKPH_2023, Coles_2016, Winick_2018} can avoid impractical measurements but rely on computationally costly and RAM-intensive convex optimization procedures. These demands are particularly challenging for HD problems, limiting the practical dimensionality to the low teens with state-of-the-art hardware~\cite{BKPH_2023}, while cutoffs \cite{Lin_2019, Kanitschar_2021} or established reduction methods \cite{Upadhyaya_2021, Kanitschar_2023, Lupo_2022, Sidhu_2024} known from DM CV-QKD, where excess dimensions are a burden that needs to be rigorously taken care of, cannot be applied for HD QKD protocols. At the same time, recent progress in high-dimensional photonics exceeds those limitations by far and brings forth the need for methods to compute secure key rates in the regime of significantly larger encoding dimensions. 

Our work addresses this issue by introducing a framework for analyzing practical high-dimensional QKD setups without relying on infeasible measurements or computationally expensive convex optimization methods. We rewrite the Devetak-Winter formula \cite{Devetak_Winter_2006} for the secure key rate as a semidefinite program (SDP) constrained by linear functions of certain observables that can be derived from the actual measurements inspired by entanglement witnesses. While the set of witness operators that are attainable via practical measurements is usually quite limited, we extend the set of possible operators by a matrix-completion argument. Instead of solving the resulting SDP directly, we derive its dual, which is guaranteed to lower bound the primal SDP, hence the secure key rate. This allows us to show that the dual problem can be simplified and rewritten such that the main remaining task is finding the largest eigenvalue of a set of parametrized matrices, followed by solving a scalar-valued optimization problem constrained by linear functions of the largest eigenvalue. When analyzing those matrices, we realized that smart choices for the witness operators make the problem fall apart into smaller subproblems, which reduces the overall dimension of the optimization problem, hence reducing the computational complexity further. While, so far, we have significantly reduced computational demands, solely relying on practically feasible measurements, the actual choice of the witness combination yielding the best key rates remains still unknown. To overcome this limitation, we modified the chosen witnesses such that they act on `orthogonal subspaces' of the quantum state, which allows us to include the choice of the relative weights of those operators directly into the optimization problem. Notably, by doing so, the witness choice can be included directly in the key rate optimization and, therefore, does not come at any additional cost. Finally, the resulting optimization problem can be solved using either Lagrange's or numerical methods. 

Our method accommodates subspace postselection, which is known to improve key rates, particularly in very noisy scenarios \cite{Doda_2021}. Finally, we illustrate our method and calculate asymptotic secure key rates for a HD temporal entanglement setup analyzed in \cite{BKPH_2023}. 

\textbf{Protocol.---}
We analyze a general high-dimensional QKD protocol, which is comprised of:  

\textbf{\textit{1.) State Generation.}} A photon source distributes entangled quantum states $\rho_{AB}$ to Alice and Bob.  

\textbf{\textit{2.) Measurement.}} Alice and Bob randomly and independently decide to measure either in their computational bases $\{A_1^x\}_{x=0}^{d-1}, \{B_1^y\}_{y=0}^{d-1}$ or in one of the test bases $\{A_2^x\}_{x=0}^{d-1}, \{B_2^y\}_{y=0}^{d-1}$ and record their outcomes in their respective registers. 

Steps 1.) and 2.) are repeated many times. 

\textbf{\textit{3.) Sifting.}} Alice and Bob use the classical authenticated channel to communicate their measurement choice to each other and may discard certain results; they also may choose to perform subspace postselection. 

\textbf{\textit{4.) Parameter Estimation.}} The communicating parties disclose some of their measurement results over the public channel to estimate the correlations between their bit-strings. 

\textbf{\textit{5.) Error Correction \& Privacy Amplification.}} Finally, on the remaining rounds, Alice and Bob perform error correction and privacy amplification to reconcile their raw keys $X$ and $Y$ and decouple them from Eve. 

\textbf{Key Rate Calculation.---} 
 As in the asymptotic regime collective i.i.d. attacks are known to be essentially optimal \cite{Renner_2008}, worst case, after protocol execution, Eve holds a purification $\rho_{ABE}$ of Alice's and Bob's shared state $\rho_{AB}$. The following does not explicitly assume that subspace postselection is performed but generally treats a QKD protocol in dimension $d$. However, according to \cite[Theorem 1]{Doda_2021}, the asymptotic key rates for the protocol including subspace postselection, employing $l$ subspaces of dimension $D$,(s.t. $d=l \times D$), is simply given by the weighted average of $l$ full space protocols of dimension $D$, $K \geq \sum_{m=0}^{l-1} P(M=m) K_m$ such that we can easily relate them to each other by replacing $d$ by $D$ and building the weighted sum. Here, $P(M=m)$ is the probability that Alice and Bob obtain an outcome in the same subspace, and $K_m$ is the key rate obtained from this subspace. The Devetak-Winter formula \cite{Devetak_Winter_2006}
$R^{\infty} = H(X|E) - H(X|Y)$ quantifies the asymptotic key rate of a QKD protocol as the difference between Eve's lack of knowledge about Alice's bit string $X$ and the amount of information Alice needs to communicate to Bob in order to reconcile their keys $X$ and $Y$. Since the second term is purely classical and can be directly calculated from Alice's and Bob's data, we focus on the first term, which can be lower bounded by the min-entropy, $H(X|E)_{\rho}\geq H_{\textrm{min}}(X|E)_{\rho}$. The latter is connected to the logarithm of Eve's average probability of guessing Alice's key string correctly, $H_{\textrm{min}}(X|E)_{\rho} = -\log_2\left(p_{\textrm{guess}}\right)$. Thus, to lower the asymptotic secure key rate, it suffices to find a way of calculating Eve's average guessing probability. We can formulate (see, for example, Ref. \cite{Doda_2021}) the average guessing probability as an optimization problem, where we maximize the probability that Eve guesses Alice's measurement outcome correctly over all purifications $\rho_{ABE}$ (which we assume to be held by Eve) of Alice's and Bob's shared state $\rho_{AB}$ and over all possible measurements $\{E^e\}_e$ Eve might carry out, constrained by physical requirements and Alice's and Bob's observations. 
\begin{equation}\label{eq:Primal}
   \begin{aligned}
    p_{\text{guess}}& = \max_{\rho_{ABE}, \{E^{\ell}\}_{\ell=0}^{d-1}} \sum_{\ell,k} \Tr{\rho_{ABE} A_1^{\ell} \otimes B_1^{k} \otimes E^{\ell}}\\
    \text{s.t.: }& \\
    & \sum_{\ell} E^{\ell} = \mathbbm{1},& \\
    &  \Tr{\rho_{ABE}} = 1,&\\
    & w_k = \Tr{\left(\hat{W}^{(e)}_k\otimes \mathbbm{1}_E\right) \rho_{ABE}},\\
    & w_j^{L} \leq \Tr{\left(\hat{W}^{(i)}_j\otimes \mathbbm{1}_E\right) \rho_{ABE}} \leq w_j^{U},\\
    & E^{\ell} \geq 0,\\
    &\rho_{ABE} \geq 0,
\end{aligned} 
\end{equation}
for $k\in \{1,..., N_{eW}\}$, $j\in \{1,..., N_{iW}\}$ and $\ell \in \{ 0,..., d-1\}$. Here, $N_{eW}$ and $N_{iW}$ represent the number of equality and inequality witnesses. While $A_1$ and $B_1$ denote Alice's and Bob's computational basis, the operators $\hat{W}^{(e)}_k$ are observables with expectations $w_k$ known with equality and $\hat{W}^{(i)}_j$ are observables where we can only find upper and/or lower bounds $w_j^U$ and $w_j^L$ for its expectations. As we derive in Appendix \ref{APDX:DerivationDual}, the dual of this SDP can be brought into the form
\begin{equation}\label{eq:Dual}
\begin{aligned}
 &\min_{\{y_k\}_{k=0}^{N_{eW}}, \{z_j^U, z_j^L\}_{j=1}^{N_{iW}}} y_0 + \sum_{k=1}^{N_{eW}} y_k w_k + \sum_{j=1}^{N_{iW}} \left(z_j^U w_j^U - z_j^L w_j^L\right) \\
   &~~~~\mathrm{s.t. }\\
   &~~~~~~~ y_0 \geq \lambda_{\text{max}}\left(M_{\ell}\right) ~~~ \forall \ell=0,...,d-1 \\
   &~~~~~~~ z_j^L, z_j^U\geq 0 ~~~ \forall j=1,...,N_{iW}\\
   &~~~~~~~y_k \in \mathbb{R} ~~~ \forall k=0,...,N_{eW},
   \end{aligned}
\end{equation}
where $\lambda_{\text{max}}(M_\ell)$ denotes the largest eigenvalue of $M_{\ell}:=\ketbra{\ell} \otimes \mathbbm{1}_d -\sum_{k=1}^{N_{eW}} y_k W^{(e)}_k - \sum_{j=1}^{N_{iW}}(z_j^U-z_j^L)W^{(i)}_j$. Due to the SDP duality theory, every solution of this dual problem is a valid upper bound for the guessing probability, giving rise to a valid lower bound on the secure key rate. By this procedure, we reduced the task of solving a computationally expensive optimization problem already computationally infeasible for medium dimensions to finding the largest eigenvalue of a matrix and solving a much simpler optimization problem. The matrix $M_{\ell}$ is a function of the observables chosen to formulate the initial optimization problem for the guessing probability. Thus, obviously, the choice of the observables heavily influences both the structure of the optimization problem, hence difficulty, and the final result, hence the obtained secure key rates. At the same time, possible choices are limited to observables accessible by the measurements Alice and Bob can carry out in their labs. For practical setups, this often will severely limit possible choices. We overcome this limitation by applying a matrix completion technique known from Refs.~\cite{Tiranov_2017, Martin_2017} (see also Appendix \ref{APDX:MatrCompl}) to $r:= \Re\left(\rho_{AB}\right)$ and obtain $|r_{j,l}| \geq \frac{r_{j,k}r_{k,l}-\sqrt{\left(r_{j,j}r_{k,k}-r_{j,k}^2\right)\left(r_{k,k}r_{l,l}-r_{k,l}^2 \right)}}{r_{k,k}}$. Starting from density matrix entries that are known from (or at least bounded by) measurements, this allows us to iteratively obtain bounds on missing entries, extending the number of possible observables. 

It remains to calculate the largest eigenvalue of $M_{\ell}$, or bounds thereof, for a particular choice of observables. Strategies for upper-bounding or calculating eigenvalues heavily depend on the exact structure of $M_{\ell}$, hence the choice of observables, and thus are highly problem specific. Certain choices allow for a direct calculation with particular matrix structure-specific formulas; others can be tackled via generalized versions of the Sherman-Morrison formula, while some choices only allow for eigenvalue bounds, which introduce looseness in the obtained key rates. For all those approaches, the optimization problem can be tackled using Lagrange's method, yielding quick results at the cost of potentially lower key rates due to tailoring the observables for analytic solvability rather than maximal rates. Alternatively, one may pursue a semi-analytic approach, choosing observables to maximize the key rate and solve for the largest eigenvalue numerically, requiring numerical optimization to solve the problem in Eq. (\ref{eq:Dual}). Since each solution of the dual problem is a valid upper bound for the guessing probability by construction, this still leads to reliable lower bounds on the secure key rates. While this method is slightly more time-consuming than the purely analytical methods outlined earlier, it nevertheless reduces computational complexity significantly, which is additionally aided by two of our main contributions. First, we introduced two kinds of observables, one operator acting on the diagonals with $i\neq j$ and witnesses that act on diagonals with $i=j$ and off-diagonal elements (see Appendix \ref{APDX:OperatorChoices} for details). This allows us to reduce each of the $d$ eigenvalue problems of dimension $d^2 \times d^2$ to $d \times d$, along with several simple, $1$-d linear constraints stemming from the $i \neq j$ lines. Second, we re-grouped the remaining witness operators such that their off-diagonals act mutually exclusive on specific off-diagonals, which allows us to automatically find good witness operators by including the choice of the relative weights directly into our optimization problem. Compared to direct numerical convex-optimization approaches, we reduce a problem with $d^2\times d^2$-dimensional objective function and $d^2\times d^2$ optimization variable to a problem with linear objective along with $d\times d$-dimensional constraints and vector-valued variables, which opens the path to high dimensions while remaining flexible.\\

\textbf{Demonstration and Results.---}  Our method works generally and independently of the physical platform, requiring only partial information about the density matrix. The only task is selecting suitable operators for the dual, which can be aided by entanglement witness theory \cite{Huber_2014, Friis_2018}. For illustration purposes, we demonstrate our method  for a typical high-dimensional temporal entanglement setup, analyzed earlier in Ref. \cite{BKPH_2023} (Protocol 1), where a source prepares a  $d$-dimensional state $\ket{\Psi_1} = \ket{\text{DD}}\otimes \frac{1}{\sqrt{d}}\sum_{k=0}^{d-1}\ket{kk}$ and  Alice and Bob either measure the Time-of-Arrival (ToA), denoted as $\textrm{TT}$, or the temporal superposition of (not necessarily) neighboring time bins (TSUP), denoted as $\textrm{SS}$. We obtain the following relations between the coincidence-click elements and density matrix elements, $\textrm{TT}(i,j) = \bra{i,j|\rho_T}\ket{i,j}$ and 
\begin{align}
    \Re\left(\mel{i,j}{\rho_T}{i-1,j-1} \right) &=  \frac{1}{4} \left(D(i,j,0, 0) - D(i,j,\frac{\pi}{2}, \frac{\pi}{2})\right),\\
    \Re\left(\mel{i,j-1}{\rho_T}{i-1,j} \right) &= \frac{1}{4} \left(D(i,j,0, 0) + D(i,j,\frac{\pi}{2}, \frac{\pi}{2})\right),
\end{align}
 where $D(i,j,\phi^A, \phi^B) $ is a quantity derived from generalized $x$- and $y$-measurements, as elaborated on in Appendix \ref{APDX:ApplHDP}, and $\rho_T$ is the temporal part of the density matrix in analyzed setup. This means by performing $x$, $y$, and $z$ measurements, we can access all diagonal elements and the real parts of certain off-diagonal elements. This can now be used to bound additional off-diagonal density matrix elements with the matrix completion technique mentioned earlier (see also Appendix \ref{APDX:MatrCompl}).

Inspired by entanglement witness theory \cite{Huber_2014, Friis_2018}, and based on the measurements performed, we choose (refer to Appendix \ref{APDX:OperatorChoices} for details)  entanglement witnesses (up to constants) for the observables $\hat{W}_0, \hat{W}_1, ..., \hat{W}_{d-1}$  with expectations $w_k:=\Tr{\rho_{AB}\hat{W}_k}$, where $\hat{W}_0$ is known with equality, while the expectations of $\hat{W}_1, ...,\hat{W}_{d-1}$ can only be upper bounded. This, finally, leads to the following optimization problem 
\begin{equation}
\begin{aligned}
       & \min \gamma + T w_0 + \sum_{k=1}^{d-1} S_k w_k \\
   \mathrm{s.t. }&\\
   & \gamma \geq \lambda_{\text{max}}\left(M_{\ell}\right) ~~~ \forall \ell=0,...,d-1 \\
   & T \in \mathbb{R}\\
   & S_k \geq 0 ~~~ \forall k=1,...,d-1.
\end{aligned}
\end{equation}

We note that our choice of observables leads to the dual variable $T$ appearing exclusively in orthogonal subspaces of $M_{\ell}$, which can be used to simplify the eigenvalue problem significantly and allows for quick evaluation using our semi-analytic approach. Therefore, we split off those subspaces where analytic solutions are known and apply numerical methods to solve a reduced problem for the remaining part of the matrices $M_{\ell}$ rather than using eigenvalue bounds that introduce looseness. When solving the optimization, we note that we do not rely on finding the minimum exactly, as every found minimum upper bounds the guessing probability by construction. In what follows, we assume an isotropic noise model, $\rho = v \ketbra{\Psi_{1}} + \frac{(1-v)}{d^2} \mathbbm{1}_{d^2}$, although we want to emphasize that this is only for demonstration purposes and we do not rely on any particular noise model. 

\begin{figure}
\includegraphics[width=0.48\textwidth]{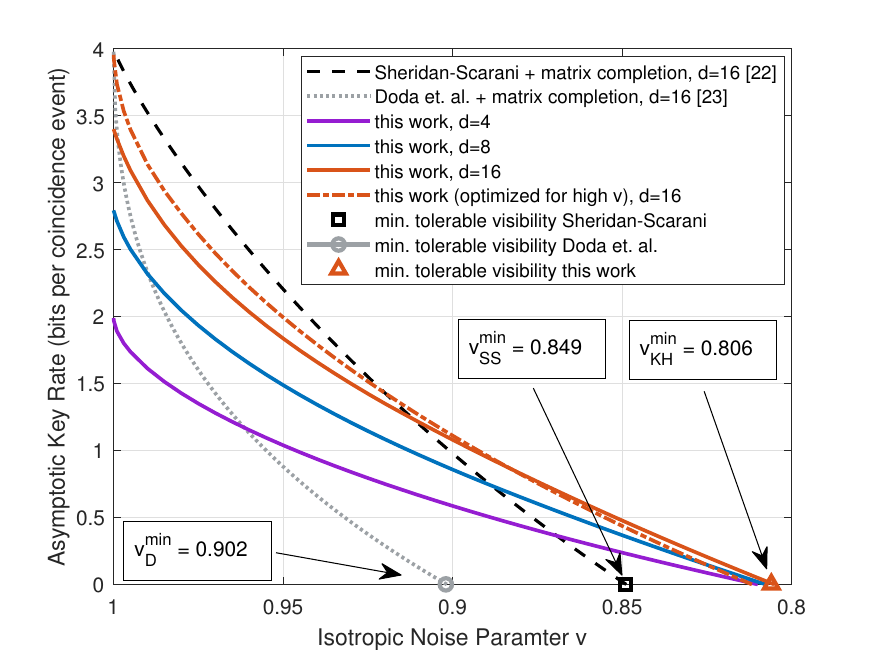}
\caption{Comparing key rates, obtained with our method with rates obtained from the methods by Doda et al. \cite{Doda_2021} (dotted grey) and Sheridan-Scarani \cite{Sheridan_2010} (dashed black) for the same data and dimension $d=16$. Our choice outperforms both in terms of maximal tolerable noise (minimal visibility $v$). Additionally, we illustrate the performance of our method for $d=4$ (purple) and $d=8$ (blue).\label{fig:Figure1}}
\end{figure}

We showcase our method for the high-dimensional QKD protocol discussed in Ref. \cite{BKPH_2023} (see also Appendix~\ref{APDX:ApplHDP}). This was motivated by rapid advancements in high-dimensional experimental photonics with measurements in platforms with overlapping measurements limited to neighbouring bins of the high dimensional discretisation (e.g. in time bin or frequency-bin approaches \cite{Martin_2017,Ponce_2022}). Beyond bin-discretisation, it would of course also work for other degrees of freedom, just our choice of dual operator is optimised for limited bin interference.

In Figure \ref{fig:Figure1}, we choose dimension $d=16$ and compare our method (solid orange) to secure key rates obtained by Ref. \cite{Doda_2021} (dotted grey) and Ref. \cite{Sheridan_2010} (dashed black) based on the same data. To ensure a fair comparison, we additionally apply matrix completion, as otherwise; one could not even calculate rates for those methods based on the available measurements. For details about the witness choices, we refer to Appendix \ref{APDX:OperatorChoices}. We show that our method outperforms both the works by Doda et. al. and Sheridan-Scarani in terms of maximal tolerable noise (minimal tolerable isotropic noise parameter $v$) significantly. While the method by Sheridan-Scarani gives slightly better key rates in the low noise regime, our technique outperforms the work by Doda et. al. in all noise regimes. Additionally, we also provide key rates for dimensions $d=4$ (solid purple) and $d=8$ (solid blue) to illustrate how high dimensions increase both the key rate and the tolerable noise further. While the operator choice for all solid curves was optimized for maximal tolerable noise, for illustration, we also provide a curve (dot-dashed orange) obtained for operators optimized for $v=0.9999$, i.e., the very low noise regime, which shows that the gap to Sheridan-Scarani in the low noise regime can be closed further and underlines that in the noiseless regime our key rates are tight. This demonstrates the flexibility of our approach, which allows choosing the operator based on the present noise level.

\begin{figure}
\includegraphics[width=0.48\textwidth]{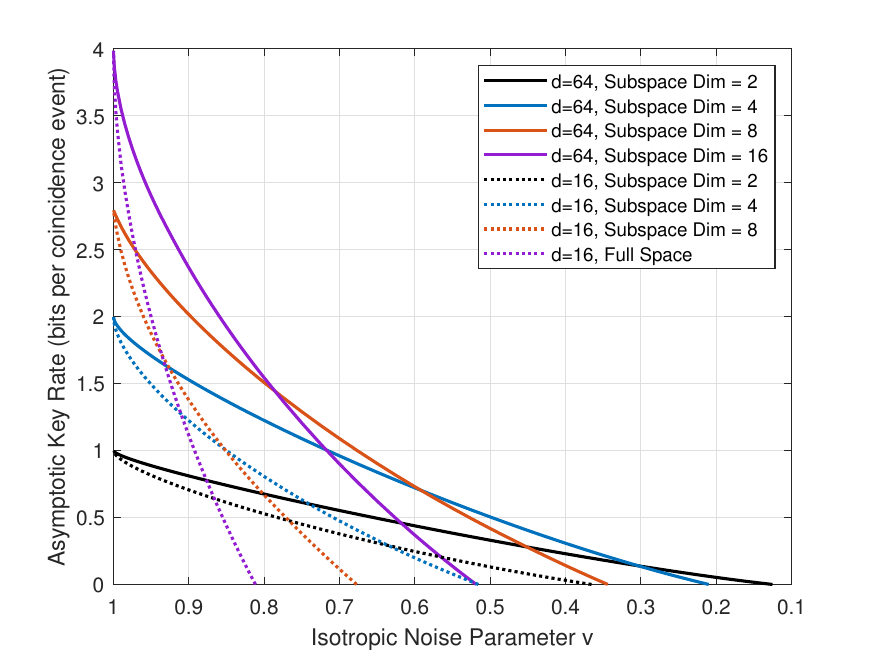}
\caption{Examination of secure key rates with applied subspace postselection for different dimensions $d$ and various subspace dimensions. \label{fig:Figure2}}
\end{figure}

Next, we combine our method with the idea of subspace postselection \cite{Doda_2021},  to harness both the improved key rate in the low-noise regime without sacrificing key generation in the high-noise regime due to excessive error correction.We illustrate our findings in Figure \ref{fig:Figure2} and observe that subspace postselection significantly enhances noise resistance and enables nonnegative key rates in high-noise regimes.

Beyond asymptotic key rates, our method extends to the finite-size regime. In our companion paper~\cite{Kanitschar_2025}, we establish finite-size security against both collective and coherent attacks in HD-QKD and show convergence to the asymptotic rates presented here. Moreover, we demonstrate the versatility of our method by establishing the security of variable-length HD-QKD protocols, which significantly enhance the expected key rates and improve the protocol's practicality.

\textbf{Discussion.---} To summarise, we set out to overcome the constraints of high-dimensional QKD by unlocking high-dimensional protocols without the need for mutually unbiased measurements or (infeasible) SDP optimization. We derive a general framework based on semi-analytic SDP duals for efficient key rate estimation and propose a family of witnesses suitable for time bin or frequency-bin photonic setups. The speed of our method scales well with increasing dimensions, which unlocks high dimensions that remain inaccessible using established purely numerical convex optimization approaches \cite{Araujo_2022, Coles_2016, Winick_2018}, meeting the demands of experimental capabilities. We show that, based on the same data, the operators employed for demonstration purposes already outperform standard techniques with additional measurement capabilities \cite{Sheridan_2010, Doda_2021} and feature key rates comparable to the full SDP in dimensions where the latter is still computationally accessible. We believe this solves a key critical issue in quantum communication, for the first time harnessing the full potential of genuinely high-dimensional entangled states in quantum key distribution with feasible measurement settings. The protocol considered for illustration is directly implementable in all higher-dimensional setups based on interfering neighboring bins, and we expect experimental demonstrations very soon. 

Additionally, in our companion paper \cite{Kanitschar_2025}, we demonstrated the finite-size security under coherent attacks for high-dimensional QKD protocols and extended those results to variable-length HD-QKD protocols, illustrating how our method provides the first full security analysis for a practical and competitive high-dimensional quantum communication protocol.

\begin{acknowledgements}
F.K. thanks Matej Pivoluska for fruitful discussions and precious feedback, and Fabien Clivaz for enlightening discussions, as well as Alexandra Bergmayr-Mann and Roman Sola\v{r} for valuable feedback on earlier versions of this work. This work has received funding from the Horizon-Europe research and innovation programme under grant agreement No 101070168 (HyperSpace).
\begin{figure}[htb!]
\centering
\includegraphics[width=0.4\columnwidth]{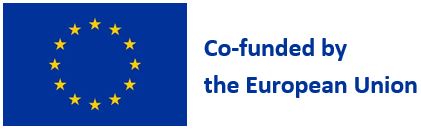}
\end{figure}
 \end{acknowledgements}
 
\appendix
\onecolumngrid
\newpage

\section{Entropic quantities used} \label{APDX:Entropies}
In this section, we define the (conditional) entropies and the related notation used in this work. For quantum mechanical state in a Hilbert space $\mathcal{H}$, described by a density operator $\rho$, the von Neumann entropy is given by 
\begin{equation*}
H(\rho):=\Tr{\rho \log_2\left(\rho\right)}.
\end{equation*}
In case the system is called $A$, represented by $\mathcal{H}_A$, we write $H(\rho_A) = H(A)_{\rho}$. This notation comes in handy in what follows. In bipartite quantum systems $\mathcal{H}_A\otimes\mathcal{H}_B$, for a state $\rho_{AB}$, we can define the conditional von Neumann entropy of $A$ given $B$ as 
\begin{equation*}
H(A|B)_{\rho} := H(AB)_{\rho} - H(B)_{\rho}.
\end{equation*}
In case the underlying variable is not a quantum state, but a random variable (i.e., diagonal), this definition naturally reduces to the Shannon entropy.

Another entropic measure used in this work is the min-entropy, which, for two systems $A$ and $B$, quantifies the maximum amount of uniform randomness that can be extracted from system $A$ given system $B$. Consider Hilbert spaces $\mathcal{H}_A$ and $\mathcal{H}_B$ and density operators $\rho_{AB}$ and $\sigma_B$ over $\mathcal{H}_A \otimes \mathcal{H}_B$ and $\mathcal{H}_B$, respectively. The min-entropy of $\rho_{AB}$ relative to $\sigma_B$ is defined as
\begin{equation*}
    H_{\textrm{min}}(\rho_{AB}||\sigma_B) := - \log_2 \inf\{\lambda \in \mathbb{R}:~ \lambda \mathbbm{1}_{A} \otimes \sigma_B \geq \rho_{AB}\}.
\end{equation*}
The min-entropy A conditioned on B of the state $\rho_{AB}$ is given by 
\begin{equation*}
H_{\text{min}}(A | B)_{\rho} := \sup_{\sigma_B \in \mathcal{D}_{\leq}(\mathcal{H}_B)} H_{\text{min}}(\rho_{AB} || \sigma_B),
\end{equation*}
where $\mathcal{D}_{\leq}(\mathcal{H}_B)$ denotes the set of all (subnormalized) density operators on $\mathcal{H}_B$. As we use in the main part of this work, the latter quantity can be related to an eavesdropper's guessing probability in quantum communication tasks.

\section{Derivation of the Dual Problem} \label{APDX:DerivationDual}
The key-rate-finding problem can be formulated as the semidefinite program given in Eq. (\ref{eq:Primal}). In what follows, we derive its dual form.

\begin{align*}
    &p_g = \max \sum_{l=0}^{d-1} \Tr{\rho_l \left(A_1^l\otimes \mathbbm{1}_d\right)}\\
     \mathrm{s.t. }&\\
    &w_j^{L} \leq \Tr{\hat{W}^{(i)}_j \sum_{l=0}^{d-1}\rho_l} \leq w_j^{U} ~~~\forall j=1,..., N_{iW}\\
    & w_k = \Tr{\hat{W}^{(e)}_k \sum_{l=0}^{d-1} \rho_l} ~~~\forall k=1,..., N_{eW}\\
    &\rho_l \geq 0\\
    & \Tr{\sum_{l=0}^{d-1} \rho_l} = 1.
\end{align*}
Here, $N_{iw}$ is the number of inequality constraints, and $N_{eW}$ is the number of equality constraints, while $w_j^{L}$ and $w_j^U$ are the lower and upper bound of the $j$-th inequality constraint. 
We rewrite this problem into
\begin{align*}
    & \max \Tr{
    \begin{pmatrix}
    \rho_0 & 0 & \hdots & 0\\
    0 & \rho_1 & \ddots & 0\\
    0 & \ddots & \ddots & 0 \\
    0 & \hdots & 0 & \rho_{d-1}
    \end{pmatrix}
    \begin{pmatrix}
        A_0^0\otimes \mathbbm{1}_d & 0 & \hdots & 0\\
        0 & A_0^1\otimes \mathbbm{1}_d & \ddots & 0\\
        0 & \ddots & \ddots & 0\\
        0 & \hdots & 0 & A_0^{d-1}\otimes \mathbbm{1}_d
    \end{pmatrix}}\\
    \mathrm{s.t. }&\\
    & w_j^L \leq \Tr{\begin{pmatrix}
    \rho_0 & 0 & \hdots & 0\\
    0 & \rho_1 & \ddots & 0\\
    0 & \ddots & \ddots & 0 \\
    0 & \hdots & 0 & \rho_{d-1}
    \end{pmatrix}
    \begin{pmatrix}
        \hat{W}^{(i)}_j & 0 & \hdots & 0\\
        0 & \hat{W}^{(i)}_j & \ddots & 0\\
        0 & \ddots & \ddots & 0\\
        0 & \hdots & 0 & \hat{W}^{(i)}_j
    \end{pmatrix}} \leq w_j^U ~~~\forall j=1,..., N_{iw}\\
    &w_k = \Tr{\begin{pmatrix}
    \rho_0 & 0 & \hdots & 0\\
    0 & \rho_1 & \ddots & 0\\
    0 & \ddots & \ddots & 0 \\
    0 & \hdots & 0 & \rho_{d-1}
    \end{pmatrix}
    \begin{pmatrix}
        \hat{W}^{(e)}_k & 0 & \hdots & 0\\
        0 & \hat{W}^{(e)}_k & \ddots & 0\\
        0 & \ddots & \ddots & 0\\
        0 & \hdots & 0 & \hat{W}^{(e)}_k
    \end{pmatrix}} ~~~\forall k=1,..., N_{eW}\\
    &1 =  \Tr{\begin{pmatrix}
    \rho_0 & 0 & \hdots & 0\\
    0 & \rho_1 & \ddots & 0\\
    0 & \ddots & \ddots & 0 \\
    0 & \hdots & 0 & \rho_{d-1}
    \end{pmatrix} \mathbbm{1}} \\
    &0 \leq \begin{pmatrix}
    \rho_0 & 0 & \hdots & 0\\
    0 & \rho_1 & \ddots & 0\\
    0 & \ddots & \ddots & 0 \\
    0 & \hdots & 0 & \rho_{d-1}
    \end{pmatrix}.
\end{align*}
Let us define some short-notation by stating the problem as
\begin{align*}
    &\max \Tr{X A}\\
    \mathrm{s.t. }& \\
    & \Tr{X \left(-\bar{W}^{(i)}_j\right) } \leq -w_j^{L} & :z_j^L \in \mathbb{R}\\
    &\Tr{X \bar{W}^{(i)}_j} \leq w_j^U & :z_j^U \in \mathbb{R}\\
    & \Tr{X \bar{W}^{(e)}_k } = w_k & y_k \in \mathbb{R}\\
    &\Tr{X \mathbbm{1}_{d^2}} = 1 & :y_0 \in \mathbb{R}\\
    & -X \leq 0 & :Z \in \mathrm{Herm}(d^2),
\end{align*}
where the variables after the $:$ indicate the dual variables associated with the constraint and $j=1,...,N_{iW}$, as well as $k=1,...,N_{eW}$. This allows us to define the following Lagrange function
\begin{align*}
    \mathcal{L}&:= \Tr{XA} + y_0 \left(1-\Tr{X \mathbbm{1}_{d^2}} \right) + \sum_{k=1}^{N_W} y_k \left(w_k - \Tr{X\bar{W}^{(e)}_k}\right) \\
    &~~+\sum_{j=1}^{N_{iw}} \left[z_j^U \left(w_j^U - \Tr{X\bar{W}^{(i)}_j}\right) + z_j^L \left(-w_j^L - \Tr{X(-\bar{W}^{(i)}_j)}\right)\right]+ \Tr{Z X}\\
    &= \Tr{\left(A-y_0\mathbbm{1} -\sum_{k=1}^{N_W} y_k \bar{W}^{(e)}_k + \sum_{j=1}^{N_{iW}}(z_j^L-z_j^U)\bar{W}^{(i)}_j +Z \right)X} + y_0 + \sum_{k=1}^{N_{eW}} y_k w_k + \sum_{j=1}^{N_{iW}} \left(z_j^U w_j^U - z_j^L w_j^L\right).
\end{align*}

Next, impose $z_j^U, z_j^L \geq 0$, $Z \geq 0$ and $A-y_0\mathbbm{1} -\sum_{k=1}^{N_W} y_k \bar{W}^{(e)}_k + \sum_{j=1}^{N_{iW}}(z_j^L-z_j^U)\bar{W}^{(i)}_j +Z =0$. Then $\mathcal{L}$ is always greater or equal to the primal objective function and independent of the primal variable $X$. With the last condition, the first trace vanishes, and we are left with the remaining terms of $\mathcal{L}$. Because of our choices, $\mathcal{L}$ is always larger or equal to the primal objective function. Therefore, we obtain the tightest upper bound by minimizing the remaining $\mathcal{L}$ over the conditions we imposed. Thus, we found the dual
\begin{align*}
   & \min y_0 + \sum_{k=1}^{N_{eW}} y_k w_k + \sum_{j=1}^{N_{iW}} \left(z_j^U w_j^U - z_j^L w_j^L\right) \\
   \mathrm{s.t. }&\\
   & A-y_0\mathbbm{1} -\sum_{k=1}^{N_{eW}} y_k \bar{W}^{(e)}_k + \sum_{j=1}^{N_{iW}}(z_j^L-z_j^U)\bar{W}^{(i)}_j +Z =0 \\
   & z_j^L, z_j^U\geq 0 ~~~ \forall j=1,...,N_{iW}\\
   &y_k \in \mathbb{R} ~~~ \forall k=0,...,N_{eW}\\
   & Z \geq 0.
\end{align*}
We can eliminate $Z$ and obtain
\begin{equation}\label{eq:genOptProb}
  \begin{aligned}
   & \min y_0 + \sum_{k=1}^{N_{eW}} y_k w_k + \sum_{j=1}^{N_{iW}} \left(z_j^U w_j^U - z_j^L w_j^L\right) \\
   \mathrm{s.t. }&\\
   & y_0\mathbbm{1} +\sum_{k=1}^{N_{eW}} y_k \bar{W}^{(e)}_k - \sum_{j=1}^{N_{iW}}(z_j^L-z_j^U)\bar{W}^{(i)}_j-A \geq 0 \\
   & z_j^L, z_j^U\geq 0 ~~~ \forall j=1,...,N_{iW}\\
   & y_k \in \mathbb{R} ~~~ \forall k=0,...,N_{eW}.
\end{aligned}   
\end{equation}

This is a semidefinite program that we hope to solve analytically. Therefore, first observe that the inequality $y_0\mathbbm{1} +\sum_{k=1}^{N_{eW}} y_k \bar{W}^{(e)}_k - \sum_{j=1}^{N_{iW}}(z_j^L-z_j^U)\bar{W}^{(i)}_j-A \geq 0$ actually means
\begin{align*}
    y_0\mathbbm{1} +\sum_{k=1}^{N_{eW}} y_k \bar{W}^{(e)}_k - \sum_{j=1}^{N_{iW}}(z_j^L-z_j^U)\bar{W}^{(i)}_j\geq \ketbra{l} \otimes\mathbbm{1}_d ~~~\forall l = 0,...,d-1,
\end{align*}
or, equivalently,
\begin{align*}
    \sum_{k=1}^{N_{eW}} y_k \bar{W}^{(e)}_k + \sum_{j=1}^{N_{iW}}(z_j^U-z_j^L)\bar{W}^{(i)}_j - \ketbra{l} \otimes \mathbbm{1}_d \geq -y_0 \mathbbm{1}_{d^2} ~~~ \forall l=0,...,d-1.
\end{align*}
In other words, all eigenvalues and therefore in particular the smallest eigenvalue of $\sum_{k=1}^{N_{eW}} y_k \bar{W}^{(e)}_k + \sum_{j=1}^{N_{iW}}(z_j^U-z_j^L)\bar{W}^{(i)}_j - \ketbra{l} \otimes \mathbbm{1}_d$ has to be larger $-y_0$ for all $l = 0,..., d-1$. To ease notation, in what follows, we call 
\begin{align*}
    M_l := \sum_{k=1}^{N_{eW}} y_k \bar{W}^{(e)}_k + \sum_{j=1}^{N_{iW}}(z_j^U-z_j^L)\bar{W}^{(i)}_j - \ketbra{l} \otimes \mathbbm{1}_d.
\end{align*}

\section{Matrix completion technique}\label{APDX:MatrCompl}
In what follows, we briefly describe the matrix completion technique by Refs. \cite{Tiranov_2017, Martin_2017} and its application to our present problem.
This technique makes use of the fact that once we have coherence between states $|i\rangle$ and $|i+1\rangle$, as well as $|i+1\rangle$ and $|i+2\rangle$, positivity of the density matrix implies a minimal coherence also between $|i\rangle$ and $|i+2\rangle$.  Since an unbalanced MZI measures the average of the first off-diagonal, this is exactly the situation we encounter here and can leverage this property iteratively to lower bound the far-off-diagonal elemenets that we have no direct measurement access to. This can be efficiently done by an SDP in the $3x3$ subspaces via an iterative method, or via a further relaxation suggested by \cite{Tiranov_2017}:
A principle minor of $M$ is $M_{I, J}$ where $I=J \subseteq \{1,...,n\}$. According to Silvester's Criterion, every Hermitian $n\times n$ matrix $M$ is positive semidefinite if every nested sequence of principle minors has a nonnegative determinant. Thus, in particular, if $M$ is hermitian and positive semidefinite, $\det\left(M_{I,J}\right)\geq 0$ holds. This can be applied to the real part of a density matrix $r = \Re(\rho) = (r_{i,j})_{i,j}$, which is positive semidefinite (as $r = \frac{1}{2}\left(\rho+\bar{\rho} \right)$) and Hermitian (since symmetric). The condition that the determinant of every proper minor is nonnegative translates to
\begin{align}
    \det\!\begin{pmatrix}
        r_{j,j} & r_{j,k} & r_{j,l} \\
        r_{k,j} & r_{k,k} & r_{k,l} \\
        r_{l,j} & r_{l,k} & r_{l,l}
    \end{pmatrix} \geq &0  ~~\Leftrightarrow  \\
   \frac{r_{j,k}r_{k,l}+\sqrt{\left(r_{j,j}r_{k,k}-r_{j,k}^2\right)\left(r_{k,k}r_{l,l}-r_{k,l}^2 \right)}}{r_{k,k}} \geq |r_{j,l}| &\geq \frac{r_{j,k}r_{k,l}-\sqrt{\left(r_{j,j}r_{k,k}-r_{j,k}^2\right)\left(r_{k,k}r_{l,l}-r_{k,l}^2 \right)}}{r_{k,k}}.
\end{align}
Note that both $r_{j,j}r_{k,k}-r_{j,k}^2\geq 0$ and $r_{k,k}r_{l,l}-r_{k,l}^2\geq 0$ since both are the determinants of a $2\times 2$ minor. Thus, we obtain the lower bound
\begin{equation}
    |r_{j,l}| \geq \frac{r_{j,k}r_{k,l}-\sqrt{\left(r_{j,j}r_{k,k}-r_{j,k}^2\right)\left(r_{k,k}r_{l,l}-r_{k,l}^2 \right)}}{r_{k,k}}.
\end{equation}
This relation can be used to iteratively derive lower bounds on the entries of $r$ as needed to bound the expectation of the chosen observable. Thus, at the price of obtaining only bounds on the expectations instead of equalities, we can significantly extend the set of accessible observables.

\section{Application to the HD Temporal Entanglement Protocol}\label{APDX:ApplHDP}
In this section, we detail the application of our method to the high-dimensional temporal entanglement protocol analyzed in Ref. \cite[Protocol 1]{BKPH_2023}, sticking close to the notation chosen there. In this protocol, a source prepares the state $\ket{\Psi_{\textrm{target}}^{\textrm{P1}}} :=  \ket{\text{DD}} \otimes \frac{1}{\sqrt{d}} \sum_{k=0}^{d-1} \ket{kk}$ which is distributed to Alice and Bob over the quantum channel. They then perform either a Time of Arrival (ToA) measurement or a Temporal Superposition (TSUP) measurement. We start by briefly summarizing the most important definitions from \cite{BKPH_2023}. The action of the temporal-superposition setup is described by 
\begin{align}
\begin{aligned}
    \hat{U} :=& \ket{\text{HH}}\!\!\bra{\text{HH}} \otimes \mathbbm{1}_T\otimes\mathbbm{1}_T + \ket{\text{HV}}\!\!\bra{\text{HV}} \otimes \mathbbm{1}_T\otimes \hat{Q}_{\phi}\hat{T} \\
    + &\ket{\text{VH}}\!\!\bra{\text{VH}} \otimes \hat{Q}_{\phi}\hat{T}\otimes\mathbbm{1}_T + \ket{\text{VV}}\!\!\bra{\text{VV}} \otimes \hat{Q}_{\phi}\hat{T}\otimes\hat{Q}_{\phi}\hat{T}.
\end{aligned}
\end{align}
The TSUP measurement is then given by $\tilde{M}_{a,b}(i,j,\phi^A, \phi^B) := \ketbra{\Psi_{a,b}(i,j,\phi^A,\phi^B)}$, where 
\begin{equation}\label{eq:DD_op}
\begin{aligned}
    \ket{\tilde{\Psi}_{1,1}(i,j,\phi^A, \phi^B)}:= &\hat{U}^{\dagger} \ket{\text{DD}, i,j} = \ket{\tilde{\Psi}_1(i,\phi^A)} \otimes \ket{\tilde{\Psi}_1(j,\phi^B)}, 
\end{aligned}
\end{equation}

\begin{equation}\label{eq:DA_op}
\begin{aligned}
    \ket{\tilde{\Psi}_{1,2}(i,j,\phi^A, \phi^B)} := &\hat{U}^{\dagger} \ket{\text{DA}, i,j} =\ket{\tilde{\Psi}_1(i,\phi^A)} \otimes \ket{\tilde{\Psi}_2(j,\phi^B)},
\end{aligned}
\end{equation}

\begin{equation}\label{eq:AD_op}
\begin{aligned}
    \ket{\tilde{\Psi}_{2,1}(i,j,\phi^A, \phi^B)} := &\hat{U}^{\dagger} \ket{\text{AD}, i,j} = \ket{\tilde{\Psi}_2(i,\phi^A)} \otimes \ket{\tilde{\Psi}_1(j,\phi^B)} ,
\end{aligned}
\end{equation}

\begin{equation}\label{eq:AA_op}
\begin{aligned}
    \ket{\tilde{\Psi}_{2,2}(i,j,\phi^A, \phi^B)} :=& \hat{U}^{\dagger} \ket{\text{AA}, i,j} =\ket{\tilde{\Psi}_2(i,\phi^A)} \otimes \ket{\tilde{\Psi}_2(j,\phi^B)},
\end{aligned}
\end{equation}
and we introduced 
\begin{equation}
 \begin{aligned}
    &\ket{\tilde{\Psi}_{x}(i,\phi)} :=\frac{1}{\sqrt{2}} \left( \ket{\text{H},i}+ (-1)^{x-1} e^{-i \phi}\ket{\text{V}, i-1} \right), \label{eq:PsiPM}
\end{aligned}   
\end{equation}
for $x \in \{1,2\}$. The ToA measurement is simply given by
\begin{equation}
    M(i,j) := \mathbbm{1}_{\text{Pol}} \otimes \ketbra{i}\otimes \ketbra{j}.
\end{equation}
We denote ToA measurement clicks by $\text{TT}(i,j)$, while TSUP clicks are denoted by $\text{SS}_{a,b}(i,j,\phi^A, \phi^B)$, where the subscript denotes which of Alice's and Bob's detectors clicked. 

It can be seen straight-forwardly that the ToA clicks directly give access to the diagonal elements of Alice's and Bob's shared density matrix $\rho_{AB}$. The action of the TSUP measurements read as follows,
\begin{align}
\label{eq:trace_1DD}
    \begin{split}
       4 \text{SS}_{1,1}(i,j,\phi^A, \phi^B) = & \Tr{\tilde{M}_{1,1}(i,j,\phi^A, \phi^B) \rho_{AB}} \\
   & =\mel{i,j}{\rho_T}{i,j} + \mel{i,j-1}{\rho_T}{i,j}e^{-\i\phi^B} \\
   &+ \mel{i-1,j}{\rho_T}{i,j} e^{-\i\phi^A} +\mel{i-1,j-1}{\rho_T}{i,j}e^{-\i(\phi^A+\phi^B)} \\
   &+ \mel{i, j}{\rho_T}{i,j-1}e^{\i\phi^B}+ \mel{i,j-1}{\rho_T}{i,j-1} \\
   &+ \mel{i-1,j}{\rho_T}{i,j-1}e^{\i(\phi^B-\phi^A)} + \mel{i-1,j-1}{\rho_T}{i,j-1} e^{-\i\phi^A}\\
   &+ \mel{i,j}{\rho_T}{i-1,j} e^{\i \phi^A} + \mel{i,j-1}{\rho_T}{i-1,j} e^{\i(\phi^A-\phi^B)} \\
   &+ \mel{i-1,j}{\rho_T}{i-1,j} + \mel{i-1,j-1}{\rho_T}{i-1,j} e^{-\i \phi^B} \\
   &+ \mel{i,j}{\rho_T}{i-1,j-1} e^{\i (\phi^A+\phi^B)} + \mel{i,j-1}{\rho_T}{i-1,j-1} e^{\i\phi^A} \\
   &+ \mel{i-1,j}{\rho_T}{i-1,j-1}e^{\i\phi^B} + \mel{i-1,j-1}{\rho_T}{i-1,j-1}
   \end{split}
\end{align}

\begin{align}
\label{eq:trace_2DD}
    \begin{split}
   4 \text{SS}_{1,2}(i,j,\phi^A, \phi^B) =& \Tr{\tilde{M}_{1,2}(i,j,\phi^A, \phi^B) \rho_{AB}}\\
   & =\mel{i,j}{\rho_T}{i,j} - \mel{i,j-1}{\rho_T}{i,j}e^{-\i\phi^B} \\
   &+ \mel{i-1,j}{\rho_T}{i,j} e^{-\i\phi^A} -\mel{i-1,j-1}{\rho_T}{i,j}e^{-\i(\phi^A+\phi^B)} \\
   &- \mel{i, j}{\rho_T}{i,j-1}e^{\i\phi^B}+ \mel{i,j-1}{\rho_T}{i,j-1} \\
   &- \mel{i-1,j}{\rho_T}{i,j-1}e^{\i(\phi^B-\phi^A)} + \mel{i-1,j-1}{\rho_T}{i,j-1} e^{-\i\phi^A}\\
   &+ \mel{i,j}{\rho_T}{i-1,j} e^{\i \phi^A} - \mel{i,j-1}{\rho_T}{i-1,j} e^{\i(\phi^A-\phi^B)} \\
   &+ \mel{i-1,´j}{\rho_T}{i-1,j} - \mel{i-1,j-1}{\rho_T}{i-1,j} e^{-\i \phi^B} \\
   &- \mel{i,j}{\rho_T}{i-1,j-1} e^{\i (\phi^A+\phi^B)} + \mel{i,j-1}{\rho_T}{i-1,j-1} e^{\i\phi^A} \\
   &- \mel{i-1,j}{\rho_T}{i-1,j-1}e^{\i\phi^B} + \mel{i-1,j-1}{\rho_T}{i-1,j-1},
   \end{split}
\end{align}

\begin{align}
\label{eq:trace_3DD}
    \begin{split}
4 \text{SS}_{2,1}(i,j,\phi^A, \phi^B) = & \Tr{\tilde{M}_{2,1}(i,j,\phi^A, \phi^B) \rho_{AB}} \\
   & =\mel{i,j}{\rho_T}{i,j} + \mel{i,j-1}{\rho_T}{i,j}e^{-\i\phi^B} \\
   &- \mel{i-1,j}{\rho_T}{i,j} e^{-\i\phi^A} -\mel{i-1,j-1}{\rho_T}{i,j}e^{-\i(\phi^A+\phi^B)} \\
   &+\mel{i, j}{\rho_T}{i,j-1}e^{\i\phi^B}+ \mel{i,j-1}{\rho_T}{i,j-1} \\
   &-\mel{i-1,j}{\rho_T}{i,j-1}e^{\i(\phi^B-\phi^A)} - \mel{i-1,j-1}{\rho_T}{i,j-1} e^{-\i\phi^A}\\
   &-\mel{i,j}{\rho_T}{i-1,j} e^{\i \phi^A} - \mel{i,j-1}{\rho_T}{i-1,j} e^{\i(\phi^A-\phi^B)} \\
   &+ \mel{i-1,j}{\rho_T}{i-1,j} + \mel{i-1,j-1}{\rho_T}{i-1,j} e^{-\i \phi^B} \\
   &-\mel{i,j}{\rho_T}{i-1,j-1} e^{\i (\phi^A+\phi^B)} - \mel{i,j-1}{\rho_T}{i-1,j-1} e^{\i\phi^A} \\
   &+\mel{i-1,j}{\rho_T}{i-1,j-1}e^{\i\phi^B} + \mel{i-1,j-1}{\rho_T}{i-1,j-1},
   \end{split}
\end{align}

\begin{align}
\label{eq:trace_4DD}
    \begin{split}
  4 \text{SS}_{2,2}(i,j,\phi^A, \phi^B) = &\Tr{\tilde{M}_{2,2}(i,j,\phi^A, \phi^B) \rho_{AB}} \\
   & =\mel{i,j}{\rho_T}{i,j} - \mel{i,j-1}{\rho_T}{i,j}e^{-\i\phi^B} \\
   &-\mel{i-1,j}{\rho_T}{i,j} e^{-\i\phi^A} +\mel{i-1,j-1}{\rho_T}{i,j}e^{-\i(\phi^A+\phi^B)} \\
   &- \mel{i, j}{\rho_T}{i,j-1}e^{\i\phi^B}+\mel{i,j-1}{\rho_T}{i,j-1} \\
   &+\mel{i-1,j}{\rho_T}{i,j-1}e^{\i(\phi^B-\phi^A)} -\mel{i-1,j-1}{\rho_T}{i,j-1} e^{-\i\phi^A}\\
   &-\mel{i,j}{\rho_T}{i-1,j} e^{\i \phi^A} + \mel{i,j-1}{\rho_T}{i-1,j} e^{\i(\phi^A-\phi^B)} \\
   &+\mel{i-1,j}{\rho_T}{i-1,j} - \mel{i-1,j-1}{\rho_T}{i-1,j} e^{-\i \phi^B} \\
   &+ \mel{i,j}{\rho_T}{i-1,j-1} e^{\i (\phi^A+\phi^B)} - \mel{i,j-1}{\rho_T}{i-1,j-1} e^{\i\phi^A} \\
   &- \mel{i-1,j}{\rho_T}{i-1,j-1}e^{\i\phi^B} + \mel{i-1,j-1}{\rho_T}{i-1,j-1}.
   \end{split}
\end{align}

Now, we combine Eqs. (\ref{eq:trace_1DD}) - (\ref{eq:trace_4DD}) in the following way and obtain 
\begin{align*}
    D(i,j,\phi^A, \phi^B) &:= \textrm{SS}_{1,1}(i,j,\phi^A, \phi^B)-\textrm{SS}_{1,2}(i,j,\phi^A, \phi^B) -\textrm{SS}_{2,1}(i,j, \phi^A, \phi^B) +\textrm{SS}_{2,2}(i,j, \phi^A, \phi^B)\\
    & = e^{-\i\left(\phi^A+\phi^B\right)} \mel{i-1,j-1}{\rho_{T}}{i,j} +e^{\i\left(\phi^A-\phi^B\right)} \mel{i,j-1}{\rho_{T}}{i-1,j} \\ &+ e^{-\i\left(\phi^A-\phi^B\right)} \mel{i-1,j}{\rho_{T}}{i,j-1} + e^{\i\left(\phi^A+\phi^B\right)} \mel{i,j}{\rho_{T}}{i-1,j-1}.
\end{align*}
Choosing $\phi_A = \phi_B = 0$ (corresponding to a generalized $x$ measurement) yields
\begin{equation}\label{eq:D00}
    D(i,j,0,0) = 2\Re\left(\mel{i,j-1}{\rho_{T}}{i-1,j}\right) +  2 \Re\left(\mel{i,j}{\rho_{T}}{i-1,j-1}\right),
\end{equation}
while for $\phi_A = \phi_B = \frac{\pi}{2}$ (corresponding to a generalized $y$ measurement) leads to
\begin{equation}\label{eq:Dpi2pi2}
    D(i,j,\frac{\pi}{2},\frac{\pi}{2}) = 2 \Re\left( \mel{i,j-1}{\rho_{T}}{i-1,j}\right) - 2\Re\left( \mel{i,j}{\rho_{T}}{i-1,j-1} \right).
\end{equation}
Adding and subtracting Eqs. (\ref{eq:D00}) and (\ref{eq:Dpi2pi2}) respectively, yields
\begin{align}
    \Re\left(\mel{i,j}{\rho_T}{i-1,j-1} \right) &=  \frac{1}{4} \left(D(i,j,0, 0) - D(i,j,\frac{\pi}{2}, \frac{\pi}{2})\right),\\
    \Re\left(\mel{i,j-1}{\rho_T}{i-1,j} \right) &= \frac{1}{4} \left(D(i,j,0, 0) + D(i,j,\frac{\pi}{2}, \frac{\pi}{2})\right).
\end{align}
Thus, performing $x$- and $y$-measurements suffices to give us access to the real parts of certain off-diagonal elements. Note that in case we do only perform a generalized $x$-measurement, we can obtain at least a bound on the real parts of those off-diagonal elements, 
\begin{equation}
  \Re\left(\mel{i,j}{\rho_T}{i-1,j-1} \right) \leq  \frac{D(i,j,0, 0)}{4} - \sqrt{\text{TT}(i-1,j) \cdot \text{TT}(i,j-1)},  
\end{equation}
which we can still use as a method.

Based on the obtained real parts, the method explained in Appendix \ref{APDX:MatrCompl}, allows us to bound additional off-diagonal elements iteratively.

\section{Operator Choices used for Demonstration\label{APDX:OperatorChoices}}

Let us start with the intuition behind possible choices of the witness for the dual. Assume we have a maximally entangled target state $|\Phi^+\rangle$, which has nonzero elements $\langle i,i|\rho|j,j\rangle$, of which we can only estimate the diagonal elements $\langle i,j|\rho|i,j\rangle$ as well as the first off-diagonal $\langle i,i|\rho|i+1,i+1\rangle$ from the neighboring bin interferences. Using matrix completion, we get lower bounds on further off-diagonal elements, albeit with smaller and smaller visibilities.

Now, let us recap the two main ideas that let us choose the observables. First, we aim to reduce the problem size of the eigenvalue problem of $d^2 \times d^2$, but in order to obtain high key rates, we do not want to drop measurements unnecessarily. Thus, we introduce one observable that leverages measurement results corresponding to diagonal elements $\langle i,j|\rho|i,j\rangle$ with $i \neq j$,
\begin{equation}
\hat{W}_0:=\sum_{\stackrel{i,j=0}{i\neq j}}^{d-1} p \ket{i,j}\!\!\bra{i,j},    
\end{equation}
where $p$ is a real number. The remaining observables are inspired by entanglement witness theory and access the remaining diagonal elements with $i = j$ and potentially all possible combinations of corresponding off-diagonals,
\begin{equation}\label{eq:Wk}
    \hat{W}_k := q_0 \sum_{i=0}^{d-1} \ket{i,i}\!\!\bra{i,i} + \frac{1}{d} \sum_{z=0}^{d-1}\sum_{i=0}^{d-1-z} q_{i,z} \left(\ket{i,i}\!\!\bra{i+z, i+z} + \ket{i+z,i+z}\!\!\bra{i, i}\right).
\end{equation}
We observe that for this choice $\hat{W}_0$ lives in a subspace orthogonal to all other operators. Even more, it allows us to split off $d^2-d$ one-dimensional subspaces, leading to separate linear constraints, leaving us back with a $d \times d$ dimensional eigenvalue problem involving $\hat{W}_k$ for $k>0$. The choice of $p$ only affects the constraint concerning $\hat{W}_0$ directly, and we choose $p = \frac{1}{d}$. In summary, this choice is the reason, why we can speed up the optimization further, as this allows us to calculate the eigenvalues of $M_{\ell}$ with contributions from $\hat{W}_0$ directly.

Second, we want to find a combination of witnesses $\hat{W}_k$, $k>0$ yielding high key rates. Now, any witness suitable for the dual must be a linear combination of the elements occurring in Eq. (\ref{eq:Wk}). The general structure of the $\hat{W}_k$ for $k>0$ was motivated by the natural entanglement witness construction for the case where the source is expected to produce $\ket{\Phi^+}$. Due to the limited measurements and after the application of matrix completion, the further off-diagonal elements are known with less accuracy, and it stands to reason that those should occur less prominently in the witness. Therefore, the best numerical values in the witness entries are unclear and expected to change with different noise and loss levels (as this changes the data quality), so the optimal choice of the $q_{z,i}$ is not obvious.  While counterintuitive at first (since the maximally entangled state is flat on the diagonal), we therefore introduce an asymmetry, motivated by the different number of known off-diagonals that connect to different diagonal elements. Thus, we choose each of the operators such that $\hat{W}_k$ only contributes in the $k$-th neighboring diagonal (choosing $q_{i,i+k}=1$, and $0$ for all other off-diagonal entries). Therefore, we chose the $\hat{W}_k$ for $k>0$ such that they have only non-zero elements in the $k$-th off-diagonal. In other words, each of the $d-1$ operators $\hat{W}_k$ acts on another off-diagonal, while they only overlap in the $z=0$ entries on the diagonal, $\ketbra{i,i}{i,i}$. 
This idea allowed us to include the choice of the optimal relative amplitude for different off-diagonals directly into the optimization problem, as, by construction, this is now the interpretation of the optimization variable $S_k$. Hence, the search for the optimal entanglement witness does not come at any additional computational cost!

While this, so far, was based on intuition, we note that the optimization confirmed the expected behavior. For high noise (low $v$), when the key rate drops to zero, only $\hat{W}_1$ contributed, which only accesses the first off-diagonal, while $S_k = 0$ for $k>1$. In contrast, with decreasing noise (increasing $v$), we observed that the number of non-zero $S_k$ started increasing, meaning that further off-diagonals started contributing.

We set all those off-diagonal entries equal to $1$, $q_{i,z} = 1$ for $z\neq 0$, which leaves us with choosing the optimal values for the non-zero diagonal elements $q_{i,0}$. We can simplify this task, as, by symmetry, we expect $q_{i,0} = q_{d-i,0}$. One could even further simplify by choosing all of them equally. Alternatively, the diagonal values can be optimized numerically.

This choice allowed us to include the choice of the relative weight of different off-diagonals in the optimization problem. The optimal relative weight will change, depending on the visibility parameter $v$. In line with intuition, for $v=1$ (so in case of no noise/loss), Bob receives the target state; thus, all weights are equal and the combined witness reassambles the maximally entangled state. While v decreases, the relative weights of far-off diagonals go down at the benefit of low-neighboring diagonals, taking the decreasing data quality for far-off diagonals into account.

In the asymptotic setting, we note that the right-hand side, $w_2:= \Tr{\rho_{AB} \hat{W}_1}$ is directly known from measurements in the computational basis, while, in contrast, $w_k:= \Tr{\rho_{AB} \hat{W}_k}$ for $k>0$ is known from the superposition measurements (for $k=1$) or can be bounded, using the completion technique introduced in Appendix \ref{APDX:MatrCompl}.

 \section{Subspace Postselection}
Our method naturally allows for subspace postselection, which we will discuss briefly in this Appendix. Intuitively, the idea behind subspace postselection is the following. According to the Devetak-Winter formula \cite{Devetak_Winter_2006}, the asymptotic secure key rate can be seen as difference between two entropic quantities, Eve's uncertainty about Alice's key string and Bob's uncertainty about Alice's key string, which quantifies the amount of error correction necessary. In the ideal case, which is no loss and no noise, the key rate potential scales logarithmically with the system dimension, as Eve's uncertainty about Alice's key scales logarithmically with the dimension of the quantum system, while due to the perfect correlations, Bob's string already aligns with Alice's without requiring error correction. As we increase noise, the error correction term scales with the dimension of the system as well, quickly trouncing the gain in key rate due to higher dimensions due to Eve's increased uncertainty. This is where subspaces of dimension $k$ (where $k$ is such that $\frac{d}{k}$ is an integer number) can be useful. Going to smaller spaces reduces both the error correction cost and Eve's uncertainty about the key, but it turns out that there are noise regimes in which the reduction of error correction cost is larger than the reduction of Eve's uncertainty, yielding overall larger keys as using the full space for the same noise level. In Figure \ref{fig:Figure2}, the optimal subspace dimension for different isotropic noise parameters can be seen by, for fixed $v$, always taking the curve with the highest key rate for fixed full-space dimension $d$. The region where the chosen subspace dimension $k$ gives the highest key rates is given by the intersection points between the curves.\\

After gaining an intuition, let us make the statement more formal. As mentioned in the main text, the asymptotic key rates for the protocol including subspace postselection, using $l$ subspaces of dimension $k$,(s.t. $d=l \times k$), is simply given by the weighted average of $l$ full space protocols of dimension $D$
\begin{equation}
K \geq \sum_{m=0}^{l-1} P(M=m) K_m.    
\end{equation}
Here, $P(M=m)$ is the probability that Alice and Bob obtain an outcome in the same subspace, and $K_m$ is the key rate obtained from subspace $m$. Thus, full-space key rates and subspace key rates can be easily related to each other by replacing $d$ by $k$ and the final key rate after applying subspace postselection is obtained by summing the resulting key rates with the probabilities $P(M=m)$ as weights.\\
This formula can be explained as follows. If Alice and Bob executed a $d$-dimensional HD-QKD protocol and decided to perform subspace postselection, one particular photon pair could have produced coincidence clicks either in one of the $l$ subspaces on both sides or produced mismatched coincidence clicks (so being detected in subspace $i$ on Alice's side, while arriving in subspace $j\neq i$ on Bob's side). Mismatched clicks (that would already have been mismatched in the full space protocol!) will be discarded, while for the matched coincidences, they proceed with the particular subspace they recorded the click. Thus, as $P(M=m)$ is the probability of obtaining a coincidence click in the same subspace, we note that $\sum_{m=0}^{l-1} P(M=m) \leq 1$, as this sum does not account for mismatched subspace clicks. However, for $v\rightarrow 1$, the mismatched clicks vanish and the sum approaches $1$. On the other side, once we decided for a subspace of dimension $k<d$, even for $v=1$, the achievable key rate is limited by $\log_2(k) < \log_2(d)$, which explains the vertical difference in key rates compared to the full-space protocol. 

\twocolumngrid
\bibliography{Bibliography}
	
\end{document}